# Gate-Level Static Approximate Adders

Subjects: Computer Science, Hardware & Architecture | Engineering, Electrical & Electronic

Padmanabhan

Submitted by:Balasubramanian

## Definition

This work compares and analyzes static approximate adders which are suitable for FPGA and ASIC type implementations. We consider many static approximate adders and evaluate their performance with respect to a digital image processing application using standard figures of merit such as peak signal to noise ratio and structural similarity index metric. We provide the error metrics of approximate adders, and the design metrics of accurate and approximate adders corresponding to FPGA and ASIC type implementations. For the FPGA implementation, we considered a Xilinx Artix-7 FPGA, and for an ASIC type implementation, we considered a 32-28 nm CMOS standard digital cell library. While the inferences from this work could serve as a useful reference to determine an optimum static approximate adder for a practical application, in particular, we found approximate adders HOAANED, HERLOA and M-HERLOA to be preferable.

## 1. Introduction

Computation-intensive technologies such as artificial intelligence, machine learning, big data and analytics, data mining, cloud computing, Internet-of-Things, etc., often deal with a data deluge, which makes processing using accurate computing techniques expensive in terms of time and resources. In such cases, it would be more feasible and economical if computing is performed such that the results are sufficiently correct, which is called approximate, inaccurate or imprecise computing.

Approximate computing encompasses hardware, software and memory storage [1][2][3]. With respect to approximate hardware, research has focused on arithmetic circuits [4] and logic circuits [5]. Within the realm of approximate arithmetic circuits, adders and multipliers have received significant attention, and this is because addition and multiplication are often performed in microprocessors [6] and digital signal processors [7].

This work discusses approximate adders, which are derived by introducing inaccuracies in an accurate adder. Basically, there are two kinds of approximate adders, namely static approximate adders (SAAs) and dynamic approximate adders (DAAs). Approximation is fixed in an SAA that may produce an accurate sum or an approximate sum corresponding to a specified accuracy in a single clock cycle and guarantees assured savings in design metrics compared to the accurate adder. On the other hand, approximation is variable in a DAA, which may produce an approximate or accurate sum on demand involving single or multiple clock cycles. Generally, DAAs comprise an additional error detection and correction logic (EDCL) to adjust their sum corresponding to a specified accuracy. While EDCL is necessary, nevertheless it represents a design overhead in DAAs. In [8]

In this work, we focus on SAAs. SAAs can be classified into three categories based on their implementation platform as: (a) suitable for FPGA implementation [9][10]; (b) suitable for ASIC type implementation [11][12][13]; and (c) suitable for both FPGA and ASIC type implementations [14][15][16][17][18][19][20][21][22][23][24][25][26][27]. With respect to ASIC type implementation, full-custom and semi-custom design approaches may be adopted. The former involves a manual transistor-level design, while the latter involves an automated gate-level design where a gate-level approximate adder can be described in a hardware description language (HDL) that can be synthesized using a logic synthesis tool. Additionally, a gate-level design is suitable for an FPGA implementation. Hence, gate-level SAAs, suitable for FPGA and ASIC type implementations, are particularly interesting since they are generic and versatile and they form the focus of this work. The objective of this work is to perform a comparative evaluation of different SAAs from the perspectives of error metrics and design metrics, and provide some inferences about which SAA(s) are better optimized. In the rest of the work, Section 2 reviews several gate-level SAAs that are suitable for FPGA and ASIC type implementations. Section 3 discusses digital image processing involving the accurate adder and various approximate adders and presents the error metrics of approximate adders. Section 4 gives FPGA- and ASIC-based design metrics of accurate and approximate adders corresponding to the application considered. Section 5 gives the concluding remarks.

## 2. Static Approximate Adders

An SAA is usually partitioned into two parts [28] viz. a precise part where addition is performed accurately and an imprecise part where addition is performed inaccurately. Less significant adder input bits are allotted to the imprecise part and more significant adder input bits are allotted to the precise part. Hence, the precise part is more significant than the imprecise part. A block schematic of the accurate adder and generic architectures of many SAAs are shown in Figure 1, where the precise and imprecise parts of the approximate adders are highlighted in blue and red, respectively.

In Figure 1, X and Y denote the adder inputs and SUM denotes the adder output. N is the adder size in bits and P is the number of input bits allotted to the imprecise part. Hence, (N–P) input bits are allotted to the precise part. If (N–P) is significantly greater than P, the speed of an approximate adder would be dictated by the speed of its precise part. Given this, for an FPGA implementation, the accurate adder and the precise part of the approximate adders can be described using the addition operator; thereby, the fast carry logic of an FPGA slice can be utilized to realize the accurate adder and approximate adders in a high-speed fashion. For a semi-custom ASIC type implementation using standard cells, the accurate adder and the precise part of the approximate adders can be described using a high-speed adder architecture such as a carry look-ahead adder (CLA), and they can be synthesized using a logic synthesis tool with speed set as the optimization goal. The precise parts of the approximate adders shown in Figures 1b–n are almost the same, except for the difference pertaining to whether the precise part may incorporate a carry input or not. Hence, the differences between various approximate adders are primarily attributed to the differences in logic between their imprecise parts.

Since the precise parts of the approximate adders can be realized in the same manner, the following discussion would deal with the imprecise parts of approximate adders shown in Figures 1b–n, which correspond to LOA, LOAWA, APPROX5, HEAA, M-HEAA, OLOCA, HOERAA, SETA, LZTA, LDCA, HOAANED, HERLOA and M-HERLOA. The approximate adders presented in [15][17] were called LOAWA and HEAA in [21], and we retain the same acronyms here for referencing. Further, the approximate adder constructed using an approximate full adder (AMA5) in [16] was called APPROX5 in [21] and we use the same acronym here for referencing. In the following discussions, OR refers to logical OR, AND (NAND) refers to logical AND (NAND), and XOR (XNOR) refers to logical XOR (XNOR) performed between Boolean literals.

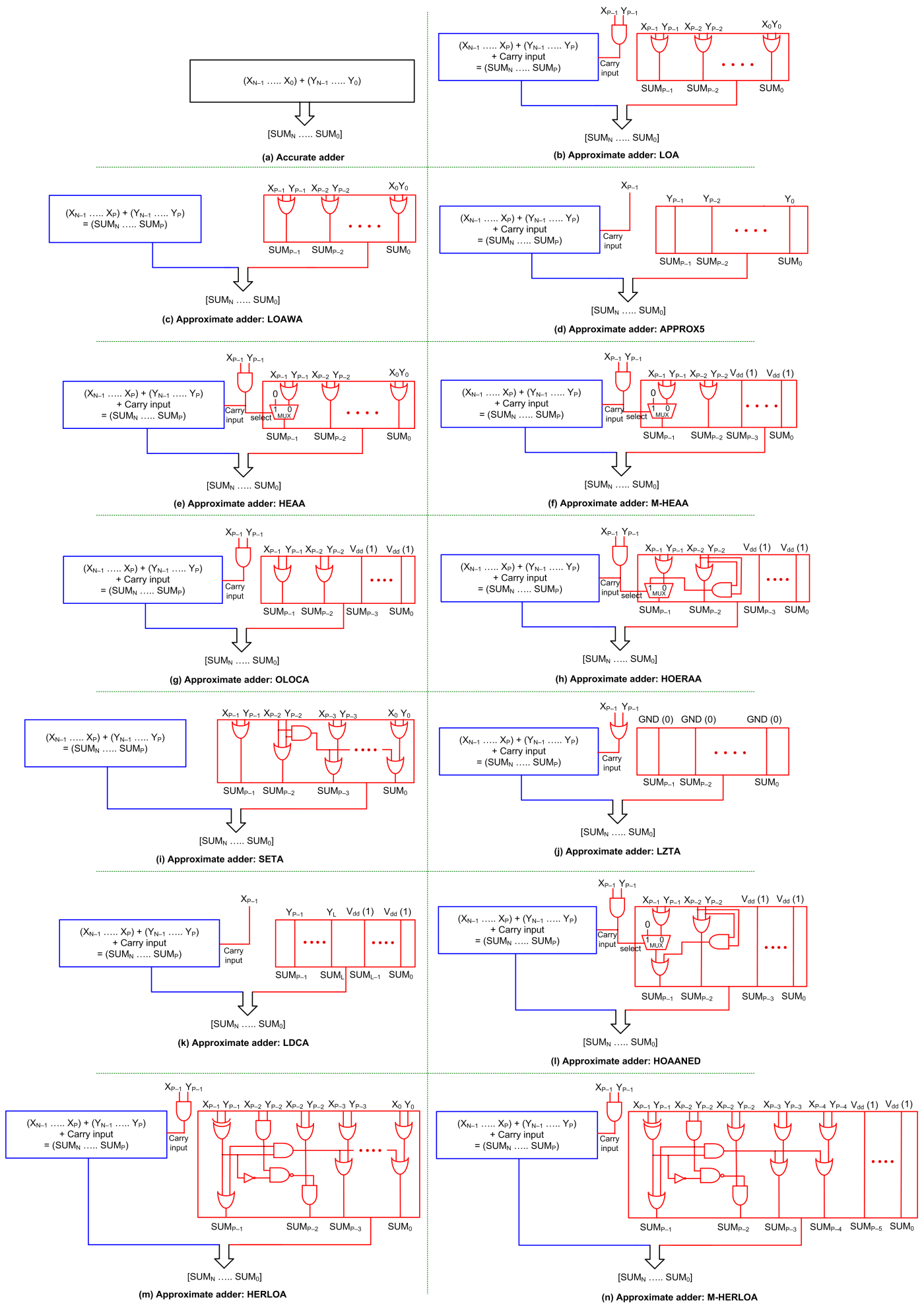

**Figure 1.** Block schematics of accurate adder and approximate adders: **(a)** Accurate adder; **(b–n)** Approximate adders.

Figure 1b shows LOA [14]. In the imprecise part of LOA, $X_{P-1}$ up to $X_0$ are bitwise OR-ed with $Y_{P-1}$ up to $Y_0$, respectively, to produce the corresponding sum bits $SUM_{P-1}$ up to $SUM_0$. $X_{P-1}$ and $Y_{P-1}$ are AND-ed to provide the carry input to the precise part.

Figure 1c shows LOAWA [15]. The logic corresponding to sum bits $SUM_{P-1}$ up to $SUM_0$ are the same for LOAWA as LOA. However, unlike LOA, there is no carry input provided from the imprecise part to the precise part in LOAWA.

In the case of APPROX5 [16], shown in Figure 1d, $Y_{P-1}$ up to $Y_0$ are forwarded as the corresponding sum bits $SUM_{P-1}$ up to $SUM_0$ using buffers, and $X_{P-2}$ up to $X_0$ are discarded. $X_{P-1}$ is given as the carry input to the precise part.

In the case of HEAA [17], shown in Figure 1e, $X_{P-2}$ up to $X_0$ are bitwise OR-ed with $Y_{P-2}$ up to $Y_0$, respectively, to produce the corresponding sum bits $SUM_{P-2}$ up to $SUM_0$. $X_{P-1}$ and $Y_{P-1}$ are AND-ed and given as the carry input to the precise part, which also serves as the select input to a 2:1 multiplexer (MUX21). If the select input of MUX21 is 0, the OR of $X_{P-1}$ and $Y_{P-1}$ is produced as $SUM_{P-1}$ and if the select input is 1, $SUM_{P-1}$ is assigned a 0.

The modified version of HEAA is shown in Figure 1f [18], which is referred to as M-HEAA in this work. The modification pertains to the assignment of a constant 1 to (P–2) least significant sum bits of the imprecise part, i.e., $SUM_{P-3}$ up to $SUM_0$. The rest of the logic of M-HEAA is the same as HEAA. Likewise, OLOCA [19], shown in Figure 1g, is a modified version of LOA in that (P–2) least significant sum bits, i.e., $SUM_{P-3}$ up to $SUM_0$ of the imprecise part of LOA are assigned a constant 1 to obtain OLOCA. Excepting for this, the rest of the logic of OLOCA is the same as LOA.

In the case of HOERAA [21], shown in Figure 1h, $SUM_{P-3}$ up to $SUM_0$ are assigned a constant 1, and $SUM_{P-2}$ is produced by OR-ing $X_{P-2}$ and $Y_{P-2}$ like M-HEAA and OLOCA. Like HEAA and M-HEAA, $X_{P-1}$ and $Y_{P-1}$ are AND-ed and given as the carry input to the precise part and also to the select input of a MUX21. If the select input of MUX21 is 0, the OR of $X_{P-1}$ and $Y_{P-1}$ is produced as $SUM_{P-1}$ and if the select input is 1, the AND of $X_{P-2}$ and $Y_{P-2}$ is produced as $SUM_{P-1}$.

In the case of SETA [22], shown in Figure 1i, the imprecise part does not supply a carry input to the precise part. The OR of $X_{P-1}$ with $Y_{P-1}$ and $X_{P-2}$ with $Y_{P-2}$ produce sum bits $SUM_{P-1}$ and $SUM_{P-2}$, respectively. The AND of $X_{P-2}$ and $Y_{P-2}$ is individually OR-ed with the respective bitwise OR-ed outputs of $X_{P-3}$ up to $X_0$ with $Y_{P-3}$ up to $Y_0$ to produce the corresponding sum bits $SUM_{P-3}$ up to $SUM_0$.

LZTA [23] is shown in Figure 1j, where all the sum bits of the imprecise part, i.e., $SUM_{P-1}$ up to $SUM_0$ are assigned a constant 0. As a result, $X_{P-2}$ up to $X_0$ and $Y_{P-2}$ up to $Y_0$ are discarded, and $X_{P-1}$ and $Y_{P-1}$ are OR-ed and given as the carry input to the precise part.

In the case of LDCA [24], shown in Figure 1k, the imprecise part is subdivided into two sections of size L bits and (P–L) bits, and these two sections are typically equal in size. The sum bits corresponding to the L bit section, i.e., $SUM_{L-1}$ up to $SUM_0$, are assigned a constant 1. In the (P–L) bit section, $Y_{P-1}$ up to $Y_L$ are forwarded as the sum bits $SUM_{P-1}$ up to $SUM_L$ through buffers, and $X_{P-1}$ is given as the carry input to the precise part.

HOAANED [25] is shown in Figure 1l. Just like M-HEAA, OLOCA and HOERAA, $SUM_{P-3}$ up to $SUM_0$ are assigned a constant 1 in HOAANED, and $X_{P-2}$ and $Y_{P-2}$ are OR-ed to produce $SUM_{P-2}$. Like HEAA, M-HEAA and HOERAA, in HOAANED, $X_{P-1}$ and $Y_{P-1}$ are AND-ed and given as the carry input to the precise part and also as the select input of a MUX21. If the MUX21 select input is 0, the OR of $X_{P-1}$ and $Y_{P-1}$ and the AND of $X_{P-2}$ and $Y_{P-2}$ are OR-ed to produce $SUM_{P-1}$; otherwise, the AND of $X_{P-2}$ and $Y_{P-2}$ alone would yield $SUM_{P-1}$.

HERLOA [26], shown in Figure 1m, consists of a unique logic in the imprecise part. $X_{P-1}$ and $Y_{P-1}$ are XOR-ed and $X_{P-2}$ and $Y_{P-2}$ are AND-ed and these two are then OR-ed to produce $SUM_{P-1}$. The XOR of $X_{P-1}$ and $Y_{P-1}$ is complemented and NAND-ed with the AND of $X_{P-2}$ and $Y_{P-2}$, which is then AND-ed with the OR of $X_{P-2}$ and $Y_{P-2}$ to produce $SUM_{P-2}$. The XOR of $X_{P-1}$ and $Y_{P-1}$ and the AND of $X_{P-2}$ and $Y_{P-2}$ are AND-ed and this is individually OR-ed with the respective bitwise OR-ed outputs of $X_{P-3}$ up to $X_0$ with $Y_{P-3}$ up to $Y_0$ to produce the corresponding sum bits $SUM_{P-3}$ up to $SUM_0$. Like LOA, HEAA, M-HEAA, OLOCA, HOERAA and HOAANED, $X_{P-1}$ and $Y_{P-1}$ are AND-ed and given as the carry input to the precise part in HERLOA.

M-HERLOA [27], shown in Figure 1n, is a modification of HERLOA in that the logic corresponding to more significant sum bits of the imprecise part (here, $SUM_{P-1}$ up to $SUM_{P-4}$) are retained the same as HERLOA and the remaining less significant sum bits of the imprecise part (here, $SUM_{P-5}$ up to $SUM_0$) are assigned a constant 1. However, the optimum number of least significant sum bits in the imprecise part, which may be assigned a constant 1 in M-HERLOA is best decided depending on which assignment enables reduced error metrics commensurate with a target application.

## 3. Digital Image Processing Using Accurate and Approximate Adders

We considered digital image processing (reconstruction) as a practical application, as in [28], to evaluate the performance of different approximate adders vis-à-vis the accurate adder. We considered many digital images with a grayscale resolution of 8 bits and a spatial resolution of 512 × 512 for experimentation. Image processing was performed as described in [25], whereby an original image was translated into a matrix form which was then processed by computing fast Fourier transform (FFT) and inverse fast Fourier transform (IFFT) accurately or approximately. The matrix output was subsequently re-translated into a digital image. Integer Fourier transforms were computed wherein multiplication was performed accurately, while addition was performed accurately or approximately. To perform accurate addition, we used the accurate adder and to perform approximate addition, we used different approximate adders individually. We considered a 32-bit addition as in [28], which implies that the size of the accurate adder and approximate adders are 32 bits. It was ensured that no data loss or overflow occurred during the FFT and IFFT computations.

Having an optimum imprecise part in an approximate circuit is important as that would pave the way for an acceptable compromise between output quality and savings in design metrics gained by an approximate circuit compared to the accurate circuit [21,25]. It was observed in [11,16] that for digital image processing and digital video encoding applications, the approximation limit may be optimally specified in the range of 7 to 9 least significant bits while considering a 32-bit arithmetic. Following a trial-and-error approach, as discussed in [25], the optimum imprecise part of the approximate adders was determined as 10 bits in size and the optimum precise part as 22 bits in size.

An example image viz. *cameraman*, which was processed accurately and approximately using accurate and approximate adders is shown in Figure 2 for an illustration. Two figures of merit viz. peak signal to noise ratio (PSNR) [29] and structural similarity index metric (SSIM) [30] were estimated to ascertain the quality of reconstructed images, and they are given above the images in Figure 2. While PSNR is a figure of merit widely used in digital signal processing, SSIM is a figure of merit of specific relevance for digital image processing. A high value of PSNR indicates less distortion in an image. SSIM is estimated by comparing a reference (original) image with a target image. Here, the target image may refer to an accurately or approximately reconstructed image. SSIM ranges from 0 to 1 decimal, with 0 indicating no similarity and 1 indicating a perfect similarity between the reference and target images. Hence, a high value of SSIM is also preferred. A perusal of Figure 2 would reveal major or minor distortions in the form of grains, spots and/or shaded regions in the images obtained using approximate adders compared to the images obtained using the accurate adder.

The image reconstructed by computing accurate FFT and IFFT involving accurate addition is shown in Figure 2a, while the images reconstructed by computing approximate FFT and IFFT involving approximate additions are shown in Figures 2b–n.

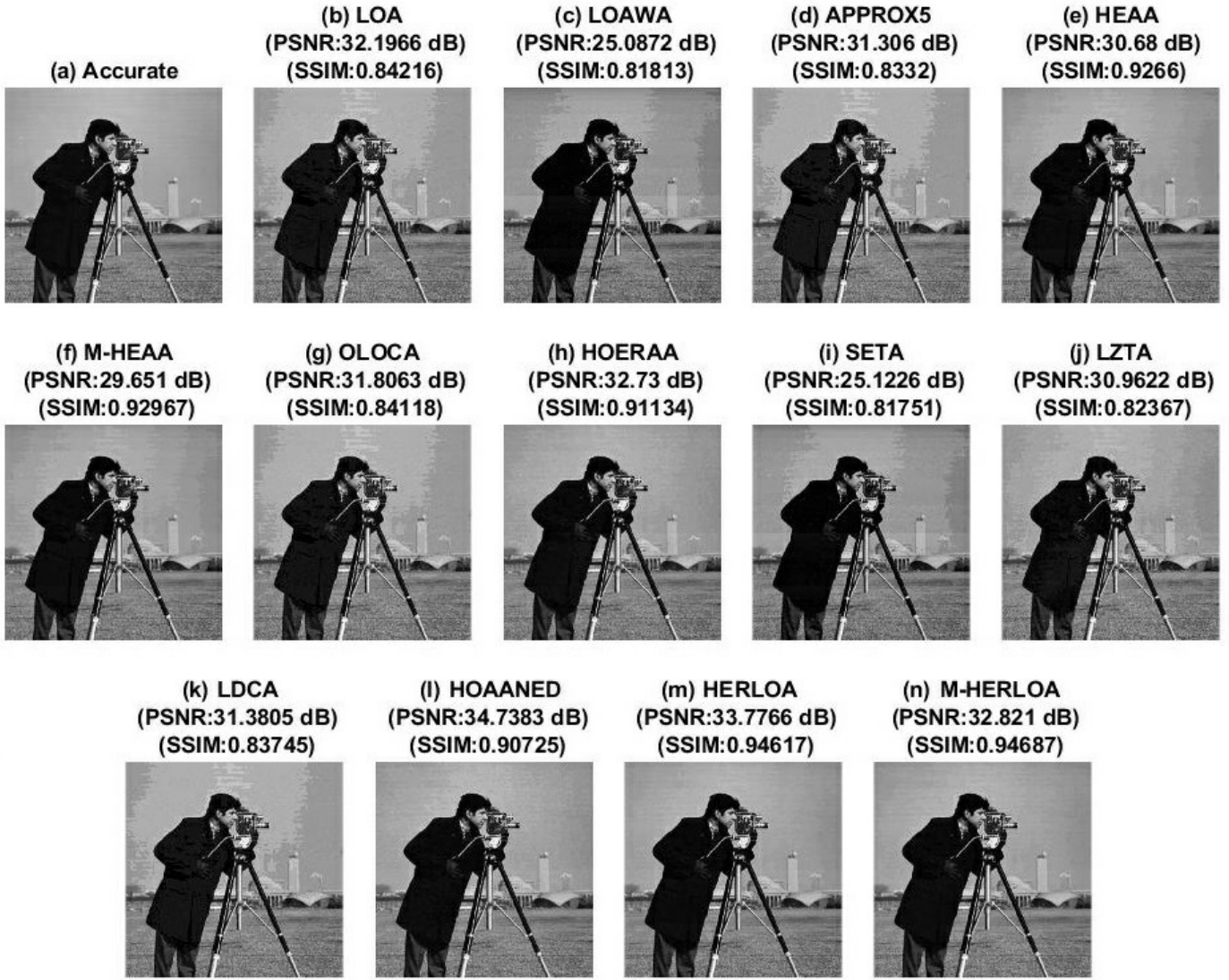

**Figure 2.** The *cameraman* image processed accurately and approximately using (**a**) accurate adder and (**b**–**n**) approximate adders.

Due to the accurate computation, PSNR = ∞ for Figure 2a and its SSIM = 1. PSNR and SSIM calculated for the images reconstructed using different approximate adders are given in Tables 1 and 2, respectively. From Figure 2 and Tables 1 and 2, it is noted that among the approximate adders, HOAANED consistently results in an improved PSNR and this is attributed to its near-normal error distribution characteristic. HOAANED also enables an enhanced SSIM in comparison with many approximate adders, except HERLOA and M-HERLOA. HERLOA and M-HERLOA consistently result in almost the same SSIM, which is greater than the SSIM of images reconstructed using other approximate adders, and this is due to a better approximate logic employed in their imprecise parts. To validate this, an error analysis was performed by supplying one million random inputs to the accurate adder and approximate adders. The extent of error occurring in the approximate adders relative to the accurate adder was plotted in the form of an error distribution, as shown in Figure 3, which portrays the error magnitudes in terms of their percentage occurrence.

Two well-known error metrics, namely mean absolute error (MAE) and root mean square error (RMSE) were calculated for the approximate adders relative to the accurate adder by considering the application of one million random input vectors. The equations for MAE and RMSE are given in [34]. MAE is also called mean error distance in the literature. Nevertheless, RMSE is more important since it better quantifies the extent of signal degradation in digital signal processing [31].

**Table 1.** PSNR (in dB) of various digital images reconstructed using different approximate adders.

| Approximate Adder | Barbara | Boat | Einstein | Lake | Cameraman | Peppers | Woman | Average PSNR |
|---|---|---|---|---|---|---|---|---|
| LOA | 32.4863 | 32.5604 | 32.5567 | 32.6313 | 32.1966 | 32.6581 | 32.8121 | 32.5574 |
| LOAWA | 25.1106 | 24.8022 | 25.7325 | 25.2703 | 25.0872 | 25.1460 | 25.2304 | 25.1970 |
| APPROX5 | 31.6881 | 31.8445 | 31.8320 | 31.7789 | 31.3060 | 31.8853 | 32.1200 | 31.7793 |
| HEAA | 30.6490 | 30.5959 | 31.0126 | 30.6447 | 30.6800 | 30.7053 | 30.8507 | 30.7340 |
| M-HEAA | 29.6692 | 29.5523 | 30.1740 | 29.6633 | 29.6510 | 29.6921 | 29.8162 | 29.7454 |
| OLOCA | 32.0496 | 32.1698 | 32.1424 | 32.1815 | 31.8063 | 32.2262 | 32.3729 | 32.1355 |
| HOERAA | 32.9709 | 33.0211 | 33.1791 | 32.9155 | 32.7300 | 33.0998 | 33.2847 | 33.0287 |
| SETA | 25.1447 | 24.8346 | 25.7657 | 25.3066 | 25.1226 | 25.1806 | 25.2653 | 25.2314 |
| LZTA | 30.8740 | 30.9092 | 31.0290 | 30.8975 | 30.9622 | 31.0619 | 30.8768 | 30.9444 |
| LDCA | 31.7570 | 31.9085 | 31.8894 | 31.8521 | 31.3805 | 31.9542 | 32.1818 | 31.8462 |
| HOAANED | 34.7582 | 34.6552 | 34.7908 | 34.7423 | 34.7383 | 34.7416 | 34.7845 | 34.7444 |
| HERLOA | 33.7722 | 33.6949 | 33.9227 | 33.7501 | 33.7766 | 33.8136 | 33.8772 | 33.8010 |
| M-HERLOA | 32.8549 | 32.7319 | 33.1088 | 32.8431 | 32.8210 | 32.8586 | 32.9572 | 32.8822 |

**Table 2.** SSIM (in decimal) of various digital images reconstructed using different approximate adders.

| Approximate Adder | Barbara | Boat | Einstein | Lake | Cameraman | Peppers | Woman | Average SSIM |
|---|---|---|---|---|---|---|---|---|
| LOA | 0.8527 | 0.8602 | 0.8440 | 0.8666 | 0.8422 | 0.8447 | 0.8150 | 0.8465 |
| LOAWA | 0.8396 | 0.8464 | 0.8198 | 0.8514 | 0.8181 | 0.8302 | 0.7884 | 0.8277 |
| APPROX5 | 0.8450 | 0.8461 | 0.8318 | 0.8537 | 0.8322 | 0.8284 | 0.8063 | 0.8348 |
| HEAA | 0.9426 | 0.9480 | 0.9370 | 0.9485 | 0.9266 | 0.9471 | 0.9174 | 0.9382 |
| M-HEAA | 0.9362 | 0.9426 | 0.9305 | 0.9426 | 0.9297 | 0.9458 | 0.9086 | 0.9337 |
| OLOCA | 0.8463 | 0.8517 | 0.8373 | 0.8587 | 0.8412 | 0.8359 | 0.8096 | 0.8401 |
| HOERAA | 0.9297 | 0.9358 | 0.9226 | 0.9394 | 0.9113 | 0.9279 | 0.9028 | 0.9242 |
| SETA | 0.8412 | 0.8475 | 0.8213 | 0.8527 | 0.8175 | 0.8319 | 0.7901 | 0.8289 |
| LZTA | 0.8290 | 0.8516 | 0.8234 | 0.8490 | 0.8237 | 0.8287 | 0.7813 | 0.8267 |
| LDCA | 0.8480 | 0.8484 | 0.8349 | 0.8562 | 0.8374 | 0.8313 | 0.8103 | 0.8381 |
| HOAANED | 0.9301 | 0.9361 | 0.9225 | 0.9372 | 0.9072 | 0.9286 | 0.9020 | 0.9234 |
| HERLOA | 0.9619 | 0.9660 | 0.9578 | 0.9663 | 0.9462 | 0.9643 | 0.9445 | 0.9581 |
| M-HERLOA | 0.9601 | 0.9640 | 0.9559 | 0.9648 | 0.9469 | 0.9637 | 0.9423 | 0.9568 |

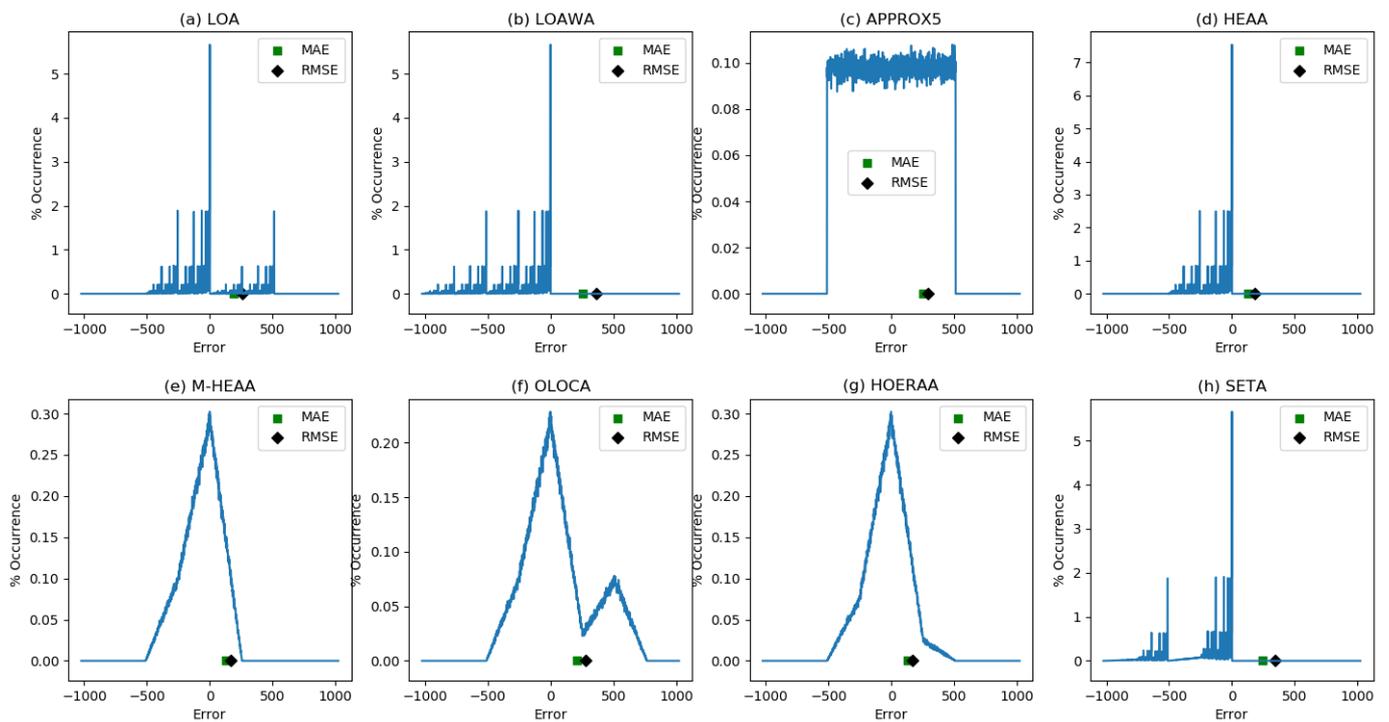

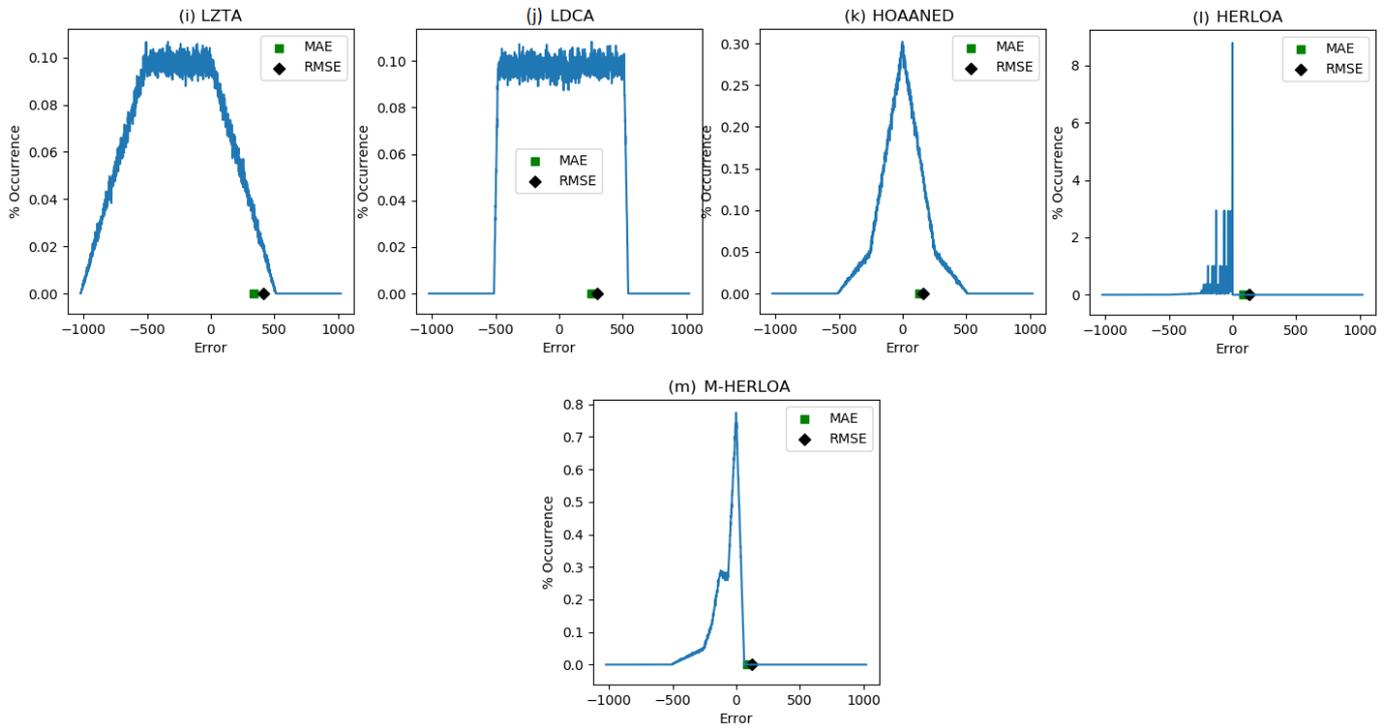

**Figure 3.** Error distribution of 32-bit approximate adders with a 10-bit imprecise part along with a highlight of their MAE and RMSE: **(a)** LOA; **(b)** LOAWA; **(c)** APPROX5; **(d)** HEAA; **(e)** M-HEAA; **(f)** OLOCA; **(g)** HOERAA; **(h)** SETA; **(i)** LZTA; **(j)** LDCA; **(k)** HOAANED; **(l)** HERLOA; **(m)** M-HERLOA. The error magnitudes are given in the X axis and the percentage of their occurrences is given in the Y axis.

From Figure 3, it is seen that HOAANED has a near-normal error distribution, which is a reflection of the fact that its positive and negative (true) error magnitudes are rather balanced and become almost neutralized on average – this is the reason for the greater PSNR of images reconstructed using HOAANED compared to the PSNR of images reconstructed using other approximate adders, as seen from Table 1.

In Figure 3, HERLOA has a restricted magnitude of error occurrences compared to the other approximate adders, and this may be the reason for the reduced distortions noticed in Figure 2m compared to Figure 2b–l. HERLOA does not have a positive error magnitude, and HERLOA is closely followed by M-HERLOA in terms of an optimized error distribution. Although the magnitude of error occurrences is relatively greater in M-HERLOA compared to HERLOA, the former has some positive error magnitudes, which contributes to an overall decrease in its MAE and RMSE.

Figure 4 depicts MAE and RMSE calculated for different approximate adders by considering the application of one million random input vectors. MAE is depicted by the blue bars and RMSE is depicted by the orange bars in Figure 4. In general, approximate adders which include a carry input in their precise part that is supplied from the imprecise part would have less errors compared to approximate adders which have disjoint precise and imprecise parts. This is because a valid carry input supplied from the imprecise part may significantly impact the output of the precise part and, thus, the overall sum. Hence, LOAWA and SETA, which do not feature an internal carry input, have higher MAE and RMSE compared to their counterparts, which feature an internal carry input. LZTA is worse since the sum bits belonging to the imprecise part of LZTA are assigned a constant 0 and so the information corresponding to the imprecise part may become completely lost during the data processing depending upon the specified inputs. Figure 4 shows that M-HERLOA has less MAE and RMSE compared to other approximate adders, with M-HERLOA having MAE and RMSE closer to HERLOA.

To achieve a higher PSNR, HOAANED is preferable and to achieve a higher SSIM, HERLOA and M-HERLOA are preferable. Nevertheless, in terms of the error metrics and image processing figures of merit combined, M-HERLOA may be preferable to its approximate counterparts.

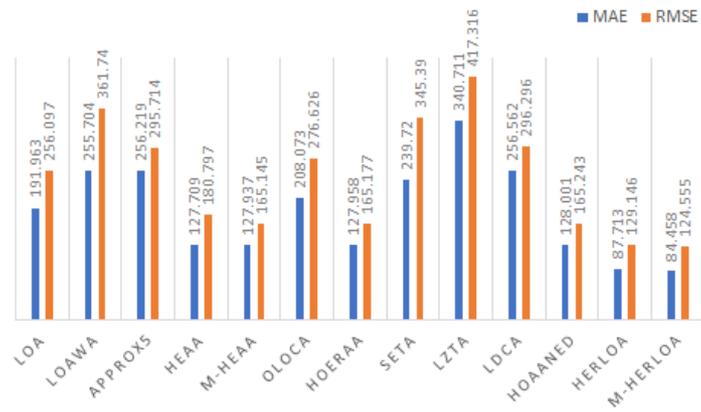

**Figure 4.** Error parameters (MAE and RMSE) calculated for different approximate adders of size 32 bits comprising a 10-bit imprecise part.

## 4. Accurate and Approximate Adders—Implementation Results

Accurate and approximate adders were implemented commensurate with the digital image processing application discussed using FPGA and ASIC design platforms. For the FPGA implementation, the accurate and approximate adders were described behaviorally in Verilog HDL and synthesized and implemented on a Xilinx Artix-7 FPGA device (part: xc7a100tcsg324-3) using Vivado design tool version: 2018.3. We described the accurate adder and the exact parts of approximate adders using the addition operator in Verilog. As a result, the fast carry logic (CARRY4) inherent in an FPGA slice was utilized to realize high speed addition. Flow_AreaOptimized_high was specified as the synthesis strategy and the default implementation strategy was used. Following an efficient FPGA design practice, a pair of register banks was provided before the adder inputs to eliminate unnecessary input–output (IO) routing delay from dominating the critical path delay. A register bank collects the adder outputs and, thus, the adder is sandwiched between the input and output register banks, with these register banks driven by a common clock. The adders were successfully synthesized and implemented, and the FPGA design metrics obtained after placement and routing namely delay (representative of minimum clock period), number of slice look-up tables (LUTs) and flip-flops consumed, and the total on-chip power consumption of the adders are given in Table 3.

**Table 3.** Design metrics of accurate and approximate adders implemented on an Artix-7 FPGA.

| Adder | Delay (ns) | LUTs | Flip-Flops | Power (W) |
|---|---|---|---|---|
| Accurate (FPGA) | 2.10 | 32 | 97 | 0.209 |
| LOA | 1.89 | 27 | 97 | 0.198 |
| LOAWA | 1.86 | 27 | 97 | 0.198 |
| APPROX5 | 1.84 | 22 | 88 | 0.200 |
| HEAA | 1.89 | 27 | 97 | 0.199 |
| M-HEAA | 1.87 | 23 | 73 | 0.188 |
| OLOCA | 1.87 | 23 | 73 | 0.187 |
| HOERAA | 1.87 | 23 | 73 | 0.188 |
| SETA | 1.85 | 31 | 97 | 0.199 |
| LZTA | 1.87 | 22 | 69 | 0.184 |
| LDCA | 1.83 | 22 | 78 | 0.195 |
| HOAANED | 1.87 | 23 | 73 | 0.188 |
| HERLOA | 1.89 | 28 | 97 | 0.199 |
| M-HERLOA | 1.90 | 25 | 79 | 0.190 |

From Table 3, we see that, in general, the approximate adders have less delay, consume fewer LUTs and flip-flops and have less on-chip power compared to the accurate FPGA adder. This is because the accurate adder is 32 bits in size, whereas the precise part of the approximate adders is only 22 bits in size, since 10 bits have been allocated to the imprecise part. Hence, the delay of the approximate adders is dominated by the delay of their precise part. Because the imprecise parts of the approximate adders have reduced logic compared to the accurate adder, fewer LUTs and/or flip-flops were required for their implementation and, thus, overall, the approximate adders require lesser resources (LUTs and flip-flops) compared to the accurate adder. For example, M-HERLOA requires 7 LUTs and 18 flip-flops less compared to the accurate FPGA adder in Table 3. Since 6 least significant sum bits were assigned a constant 1 in M-HERLOA, 12 input flip-flops and 6 output flip-flops were not required, thus saving 18 flip-flops compared to the accurate adder. Additionally, the reduction in logic of the approximate adders results in their reduced power consumption compared to the accurate adder. The differences between the resource utilization and power consumption of approximate adders are due to the differences between their imprecise part logic. The delay is almost the same for the approximate adders and only minor variations are observed between them. This is partly because the precise part of some approximate adders accepts a carry input from the imprecise part, while this is absent in the other approximate adders, and partly due to the area optimized place and route as performed by the FPGA design tool.

In Section 3, in terms of error metrics and/or image processing results, it was noted that HOAANED, HERLOA and M-HERLOA are preferable. From Table 3, it is noted that compared to the accurate FPGA adder, HOAANED has 11% less delay, requires 28.1% fewer LUTs and 24.7% fewer flip-flops, and consumes 10% less power; HERLOA has 10% less delay, requires 12.5% fewer LUTs and consumes 4.8% less power; and M-HERLOA has 9.5% less delay, requires 21.9% fewer LUTs and 18.6% fewer flip-flops, and consumes 9.1% less power.

For an ASIC type standard cell-based implementation, the accurate and approximate adders were described structurally in Verilog HDL. To realize the accurate and approximate adders for high speed, the accurate adder and precise parts of the approximate adders were described using a high speed CLA architecture [32]. The 32-bit accurate adder was described using eight 4-input CLAs, and the 22-bit precise parts of the approximate adders were described using five 4-bit CLAs and a 2-bit CLA. The 2-bit CLA may or may not include a carry input and this depends on the approximate adder architecture considered, i.e., whether the approximate adder may or may not have a carry input supplied from the imprecise part to the precise part. It may be recalled from Section 2 that LOAWA and SETA do not feature an internal carry input from the imprecise part to the precise part, while the rest of the approximate adders do.

The accurate and approximate adders were synthesized for high-speed using Synopsys Design Compiler with speed set as the optimization goal and their total area (cells area plus interconnect area) was estimated. A 32/28 nm CMOS standard cell library [33] was used for the implementation. A typical case library specification with a supply voltage of 1.05 V and an operating junction temperature of 25 °C was considered. After synthesis, the adders were simulated and their functionality was verified. Subsequently, the switching activity data obtained was used to estimate the total average power dissipation using PrimePower. PrimeTime was used to estimate the critical path delay. The adder outputs were assigned a fanout-of-4 drive strength and default wire loads were included. The ASIC-based design metrics are given in Table 4.

In Table 4, we see that all the approximate adders have the same delay and this is because their precise parts were realized for high-speed using a common CLA architecture. The areas of approximate adders, however, differ and this is due to the differences in the logic composition of their imprecise parts. Consequently, their power dissipation also differs. To assign a constant 1 to some least significant sum bits in M-HEAA, OLOCA, HOERAA, LDCA, HOAANED and M-HERLOA, tie-to-high (TIEH) standard cells were used and to assign a constant 0 to some least significant sum bits in LZTA, tie-to-low (TIEL) standard cells were used. TIEH and TIEL standard cells of [42] have the same design attributes. Given that HOAANED, HERLOA and M-HERLOA are preferable, from Table 4, it is noted that HOAANED, HERLOA and M-HERLOA have 17.9% less delay compared to the accurate CLA. Further, compared to the accurate CLA, HOAANED occupies 24.7% less area and dissipates 28.2% less power, HERLOA occupies 21.5% less area and dissipates 21.5% less power, and M-HERLOA occupies 23.1% less area and dissipates 26.7% less power.

Power-delay product (PDP), which is representative of energy and considered as a low power figure of merit, was calculated for accurate and approximate adders corresponding to FPGA and ASIC type implementations and normalized, which is shown in Figure 5. To normalize the PDP, the highest PDP corresponding to an adder (i.e., accurate adder) was considered as the baseline and this was used to divide the PDP of all the adders corresponding to FPGA and ASIC type implementations separately. The green and blue bars shown in Figure 5 represent the normalized PDP corresponding to FPGA and ASIC type implementations, respectively. Power and delay are preferred to be less for a digital design and, hence, PDP is also preferred to be less. In Figure 5, the approximate adders are found to have less PDP compared to the accurate adder, meaning the former are more energy efficient than the latter.

**Table 4.** Design metrics of accurate and approximate adders synthesized using a 32/28 nm CMOS standard digital cell library.

| Adder | Delay (ns) | Area (µm$^2$) | Power (µW) |
|---|---|---|---|
| Accurate (CLA) | 1.17 | 564.60 | 94.33 |
| LOA | 0.96 | 428.36 | 71.77 |
| LOAWA | 0.96 | 413.37 | 68.86 |
| APPROX5 | 0.96 | 424.58 | 73.54 |
| HEAA | 0.96 | 430.65 | 71.49 |
| M-HEAA | 0.96 | 422.32 | 66.11 |
| OLOCA | 0.96 | 420.03 | 66.38 |
| HOERAA | 0.96 | 430.38 | 68.82 |
| SETA | 0.96 | 419.68 | 72.94 |
| LZTA | 0.96 | 415.56 | 63.14 |
| LDCA | 0.96 | 420.07 | 68.05 |
| HOAANED | 0.96 | 425.36 | 67.73 |
| HERLOA | 0.96 | 443.28 | 74.01 |
| M-HERLOA | 0.96 | 433.94 | 69.11 |

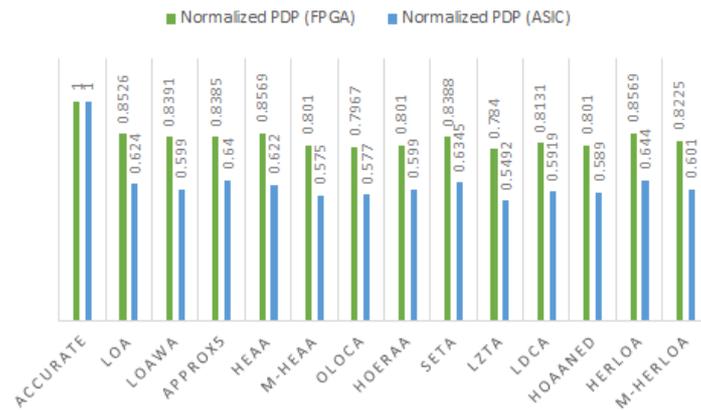

**Figure 5.** Normalized PDP of accurate and approximate adders corresponding to FPGA and ASIC type implementations.

The normalized PDP plots of the adders corresponding to FPGA and ASIC type implementations indicate a similar trend. Among the adders, LZTA is very energy efficient. However, the image processing results shown in Figure 2 and Tables 1 and 2, and the error distribution and error metrics given in Figures 3 and 4, clearly show that LZTA is not preferable. In approximate computation, output quality assumes higher precedence than savings in design metrics gained compared to accurate computation. Given this, LZTA is not preferable, although it may have a high energy efficiency. On the contrary, HOAANED, which enables a higher PSNR, and HERLOA/M-HERLOA, which enable a higher SSIM, are preferred and they report a significant improvement in energy efficiency compared to the accurate adder. From Figure 5, we note that HOAANED, HERLOA and M-HERLOA achieve 19.9%, 14.3% and 17.5% reduction in PDP, respectively, compared to the accurate adder for an FPGA implementation, and 41.1%, 35.6% and 39.9% reduction in PDP, respectively, compared to the accurate CLA for an ASIC-type implementation.

## 5. Conclusion

A comparative analysis of different gate-level SAAs, suitable for both FPGA and ASIC type implementations, has been performed in this work. Digital image processing was considered as an example application and the image processing results were shown. The error metrics of approximate adders corresponding to the image processing application were calculated and provided for a comparison. Further, the design metrics of accurate and approximate adders commensurate with the target application were provided corresponding to FPGA and ASIC type implementations. The objective is to identify those approximate adders that would facilitate an acceptable compromise between output quality and savings in design metrics compared to the accurate adder. In this context, approximate adders HOAANED, HERLOA and M-HERLOA are found to be preferable. Nevertheless, the optimum approximate adder suitable for a target application may be best determined based on a trial-and-error experimentation.[1][2][3][4][5][6][7][8][9][10][11][12][13][14][15][16][17][18][19][20][21][22][23][24][25][26][27][28][29][30][31][32][33][34]

## Keywords

approximate computing;approximate adder;digital circuits;logic design;FPGA;ASIC;VLSI design;electronics;computer engineering

Retrieved from https://encyclopedia.pub/18160